\documentclass[twocolumn,showpacs,preprintnumbers,amsmath,amssymb,superscriptaddress]{revtex4}
\usepackage{graphicx}
\usepackage{dcolumn}
\usepackage{bm}

\begin{document}
\title{Interactions between directly and parametrically driven vibration modes in a micromechanical resonator}
\author{H. J. R. Westra} \email{h.j.r.westra@tudelft.nl}
\affiliation{Kavli Institute of Nanoscience, Delft University of Technology, Lorentzweg 1, 2628CJ Delft, The Netherlands}
\author{D. M. Karabacak}
\author{S. H. Brongersma}
\author{M. Crego-Calama}
\affiliation{Holst Centre / imec the Netherlands,  High Tech Campus 31, 5656 AE, Eindhoven,  The Netherlands}
\author{H. S. J. van der Zant}
\author{W. J. Venstra} 
\affiliation{Kavli Institute of Nanoscience, Delft University of Technology, Lorentzweg 1, 2628CJ Delft, The Netherlands}

\date{\today}
\begin{abstract}
The interactions between parametrically and directly driven vibration modes of a clamped-clamped beam resonator are studied. An integrated piezoelectric transducer is used for direct and parametric excitation. First, the parametric amplification and oscillation of a single mode are analyzed by the power and phase dependence below and above the threshold for parametric oscillation. Then, the motion of a parametrically driven mode is detected by the induced change in resonance frequency in another mode of the same resonator. The resonance frequency shift is the result of the nonlinear coupling between the modes by the displacement-induced tension in the beam. These nonlinear modal interactions result in the quadratic relation between the resonance frequency of one mode and the amplitude of another mode. The amplitude of a parametrically oscillating mode depends on the square root of the pump frequency. Combining these dependencies yields a linear relation between the resonance frequency of the directly driven mode and the frequency of the parametrically oscillating mode.
\end{abstract}

\maketitle

\section{introduction}

Parametric amplification and oscillations occur when in a resonant system, one of the system parameters (e.g. spring constant, effective mass) is modulated. The principle is used in low-noise electronic amplifiers~\cite{Yurke1988:p9, Radeka1966} and to increase the broadband gain in fiber optics~\cite{Zhang:2010p023816,Marhic:1996p576, Savchenkov2007:p157}. In mechanical resonators, parametric oscillations are typically obtained by modulation of the spring constant~\cite{Lifshitz:2003p7606, Karabalin:2011p8428, Turner:1998p149, Zalalutdinov:2003p3281}. Applications of parametric resonances in nano- and micro electromechanics~\cite{Ekinci2005:p061101} (NEMS and MEMS) include quality (Q-)factor enhancement~\cite{Mahboob:2008p253109,Tamayo:2005p044903} and bit storage and bit flips using the bistable phase in a parametric oscillator~\cite{Mahboob:2008p275,Mahboob2011:p198}. Parametric amplification can also be used for noise-squeezing in a coupled qubit-resonator system~\cite{Suh:201p3990} and was recently observed in carbon nanotube resonators~\cite{Eichler:2011}.\\
\indent \indent Another interesting phenomenon in NEMS is the interaction between different vibration modes. Motivated by the trend towards large scale integration of resonators, researchers study the interactions between several resonators~\cite{Karabalin:2009p165309}. Recently, nonlinear modal interactions between two flexural modes in a clamped-clamped beam resonator~\cite{Westra:2010p7074,Dunn:2010p123109, Mahboob2011:p113411} and a cantilever~\cite{Venstra:2011} have been reported: it has been shown that the resonance frequency of one mode depends quadratically on the amplitude of another mode. \\
\indent \indent Here, we explore the modal interactions between a directly and a parametrically driven mode, yielding a \emph{linear} dependence of the resonance frequency of the directly driven mode on the pump frequency of the parametrically driven mode. In section II, the experimental conditions are provided. The following section reports on a detailed analysis of the piezoelectrical parametric amplification of a single mode. Section IV discusses the modal interactions between a directly driven and a parametrically pumped mode, and this is the central result of this work. \\

\section{Device details}
\begin{figure}[!h!t]
\includegraphics[width=85mm]{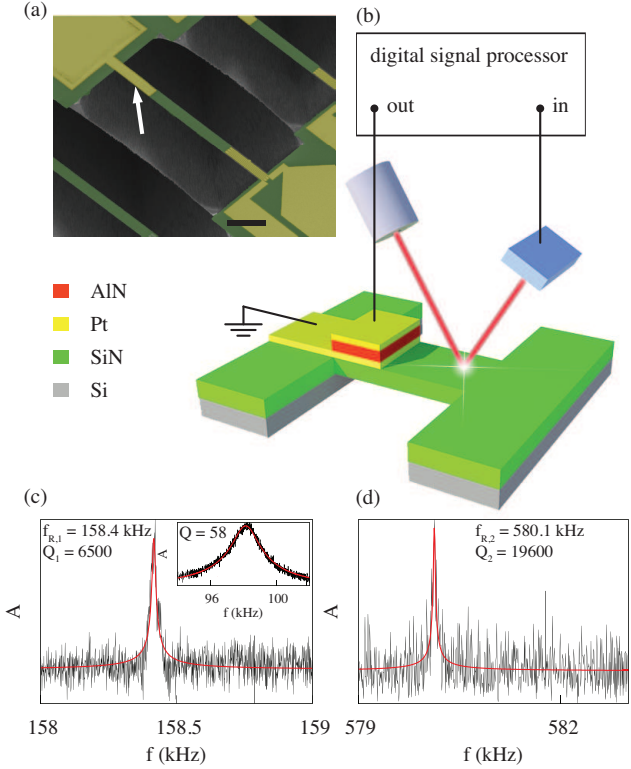}
\caption{Measurement setup. (a) False-colored scanning electron micrograph of a SiN beam with the piezo actuator (white arrow) on top (scale bar is 20 $\mu$m). (b) An optical deflection setup is used to detect motion in air and vacuum. The piezo-active AlN layer is depicted in red. The piezo actuator and photo diode are connected to a digital signal processor. (c,d) Typical frequency responses of the first (c) and second (d) mode (amplitude $A$) in vacuum. The inset in (c) shows the frequency response in air, of the resonator with length 750 $\mu$m, with a resonance frequency of 98 kHz. The response of a damped-driven harmonic oscillator is fitted through the responses to obtain Q-factors and resonance frequencies.}
\label{fig1}
\end{figure}

The resonators are clamped-clamped beams fabricated from 500 nm thick low-stress silicon nitride (SiN). A stack of platinum (Pt), aluminum nitride (AlN) and Pt (100-400-100 nm thick) is sputtered on top, to form an integrated piezoelectric transducer. Fig.~\ref{fig1}(a) shows a scanning electron micrograph of the device, the white arrow indicates the transducer. The resonators are freely suspended by a through-the-wafer etch. Two lengths are used: $L = $ 500  and 750 $\mu$m. The width of both resonators is 45 $\mu$m. Details of the fabrication procedure are described in Ref.~\cite{Karabacak:2010p7451}. An ac voltage on the piezo produces a force on the resonator and at the same time modulates its spring constant. Both the force and the spring constant depend linearly on the voltage. The voltage on the piezo, $V_\mathrm{out}$, is composed of two frequencies, one to directly excite the resonator and one to parametrically pump it, i.e. $V_\mathrm{out} = V_{\mathrm{direct}} \cos(\Omega t) + V_{\mathrm{pump}} \cos(2 \Omega t + \phi)$, where $\Omega$ is the drive frequency and $\phi$ the phase difference between the two voltages. \\

\indent \indent The motion of the resonator is measured using an optical deflection setup, as depicted in Fig.~\ref{fig1}(b).  Frequency spectrum and network analyzer measurements are implemented in a digital signal processor. Measurements are conducted in vacuum at a pressure of $10^{-4}$ mbar and at atmospheric pressure. For direct driving, the frequency responses at the first mode and second mode in vacuum are shown in Fig.~\ref{fig1}(c) and (d), with $Q_1=6500$ and $Q_2=19600$~\cite{footnote}.\\

\begin{figure}[!h!t]
\includegraphics[width=85mm]{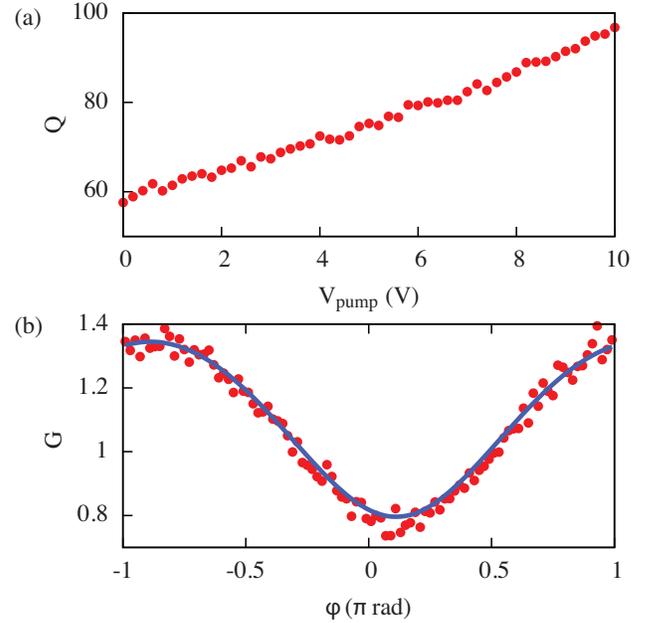}
\caption{Characterization of the parametric amplification in air. (a) The Q-factor enhancement is proportional to the parametric pump voltage. (b) Measured gain-phase relation; the blue line represents Eq. 2, with fit parameter $k_p/k_t = 0.26$.}
\label{fig2}
\end{figure}

\section{Parametric amplification of a single mode}
\indent \indent The time-dependent part of the equation of motion of the piezoelectric resonator including parametric modulation of the spring constant is described by:

\begin{equation}
m\ddot{u} + \frac{m\omega_\mathrm{R}}{Q}\dot{u}+ [m\omega_ \mathrm{R}^2 + k_p \sin(2\Omega t  + \phi)]u + \alpha u^3= F\cos(\Omega t).
\label{eq1}
\end{equation}

Here, $u(t)$ is the amplitude of the mode, $m$ is the effective mass and $F$ the direct drive force, and $\omega_{\mathrm{R}}$ is the resonance frequency. The dots denote taking the derivative to time. The spring constant is modulated at twice the drive frequency $\Omega$ with modulation strength $k_p$. $\alpha$ accounts for the Duffing nonlinearity with $\alpha > 0$ for clamped-clamped beams~\cite{Zhang:2002p139}. The parametric gain $G$ is defined by the ratio between the amplitude of the motion with and without parametric drive, and can be calculated from Eq.~\ref{eq1}~\cite{LifshitzCross,Rugar:1991p8}:
\begin{equation}
G(\phi) = \sqrt{ \frac{\cos^2 (\phi/2)}{(1 + k_p/k_t)^2} + \frac{\sin^2 (\phi/2)}{(1 - k_p/k_t)^2} }.
\label{eq2}
\end{equation} 
This equation holds for small amplitude vibrations, where the nonlinearity can be neglected. Depending on $\phi$, the motion is amplified ($G>1$) or attenuated ($G<1$). Above the threshold parametric pump, $k_p > k_t$ with $k_t = 2m\omega_\mathrm{R}^2/Q$ , the resonator is parametrically oscillating. \\
\indent \indent Parametric behavior is demonstrated for a resonator with length 750 $\mu$m vibrating in air, with $f_\mathrm{R,1} = 98$ kHz and $Q_1=58$ (frequency response in the inset of Fig~\ref{fig1}c). To amplify the motion, the resonator is driven parametrically at $2f_\mathrm{R,1}$ with $\phi = -0.75 \pi$. Figure 2(a) shows the Q-factor of the resonator as a function of the parametric pump voltage. The Q-factor increases by a factor of 1.7 when the parametric pump is 10 V. Furthermore, the phase dependence of the gain at 10 V parametric pump is plotted in Fig.~\ref{fig2}(b). The gain varies periodically with the phase difference with a period of $2 \pi$. The minimum gain is smaller than one, indicating destructive interference by an out-of-phase parametric signal. Eq.~\ref{eq2} fits the measured data well with $k_p= 0.26\, k_t$. In these experiments the parametric driving is below the parametric threshold $k_t$. A further increase of the pump voltage is not possible as this would damage the piezo-stack. To study parametric oscillation, further experiments are conducted in vacuum. Here the Q-factor improves by two orders of magnitude (Fig.~\ref{fig1}(c)), enabling post-threshold driving.\\

\begin{figure}[!h!t]
\includegraphics[width=85mm]{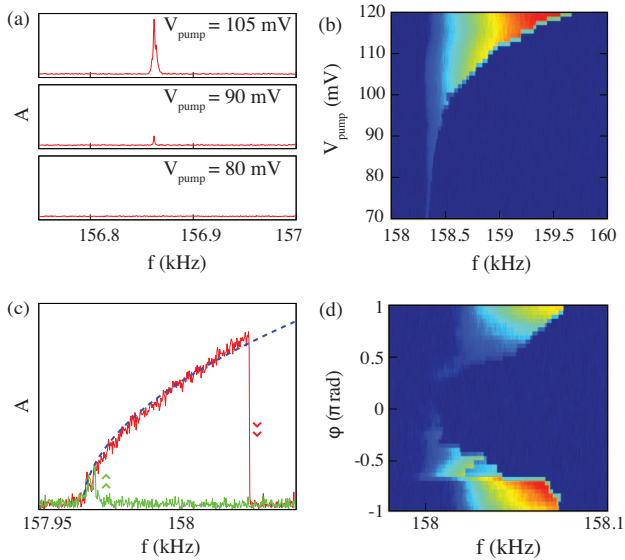}
\caption{Parametric oscillations of the first flexural mode in vacuum. (a) Frequency spectra at three pump voltages, the parametric oscillation becomes visible when $V_{\mathrm{pump}} > $ 85 mV. (b) Parametric tongue, showing frequency responses when the resonator is driven directly ($V_{\mathrm{direct}} = 5$ mV) and parametrically past the instability threshold. Color indicates the amplitude of oscillation. (c) The hysteresis between the forward (red) and reverse sweep (green) when driving parametrically ($V_{\mathrm{pump}} = 95$ mV). The blue dashed line shows the square root dependence of the amplitude ($A$) on the frequency $f$. (d) The phase dependence of the parametric oscillations at $V_{\mathrm{pump}} =$ 95 mV. The color indicates the amplitude of oscillation.}
\label{fig3}
\end{figure}

\indent \indent Figure~\ref{fig3} summarizes the measurements of the parametric oscillations performed in vacuum. A $500\, \mu$m long resonator is used, for which the frequency response is plotted in Fig.~\ref{fig1}(c). Frequency spectra are measured for three parametric pump voltages in Fig.~\ref{fig3}(a). At 80 mV no sign of oscillation is observed (lower panel), and the onset of parametric oscillation is found around 85 mV as shown in the middle panel. A further increase of the pump voltage (upper panel) results in a larger oscillation amplitude. Here, the nonlinear term in Eq.~\ref{eq1} results in an amplitude-dependent resonance frequency. Fig~\ref{fig3}(b) shows network analyzer measurements of the resonator amplitude (color scale) as a function of the pump voltage. The resonator is driven directly and parametrically with $\phi = -0.65 \pi$. A direct drive signal, weak enough to operate the resonator in the linear regime when $V_{\mathrm{pump}}=0$, is applied to initiate the motion. The motion of the weakly driven resonator is coherently amplified by the parametric excitation and the amplitude increases with $V_{\mathrm{pump}}$. The observed frequency stiffening is expected for a cubic spring constant $\alpha>0$. The oscillation sustains over a few kHz when the frequency is swept forward. The amplitude shows a hysteretic response when the frequency is swept back, see Fig.~\ref{fig3}(c). The amplitude of the oscillation depends on the square root of the frequency (dashed blue line)~\cite{LifshitzCross}. To study the relation between the parametric oscillation amplitude and the phase $\phi$, the resonator is parametrically excited above the threshold. Fig.~\ref{fig3}(d) shows the amplitude of the oscillation when the direct drive and frequency is swept while varying the phase difference. Depending on the phase between the direct initiator drive and the parametric excitation, constructive or destructive interference occurs which results in amplification or attenuation of the motion induced by the initiator signal. The maximum parametric amplification is found at a phase difference of $-\pi$ and $\pi$. The experiments described above clearly demonstrate the parametric behavior.\\

\section{Coupling between parametric and direct driven modes}
\begin{figure}[!h!t]
\includegraphics[width=85mm]{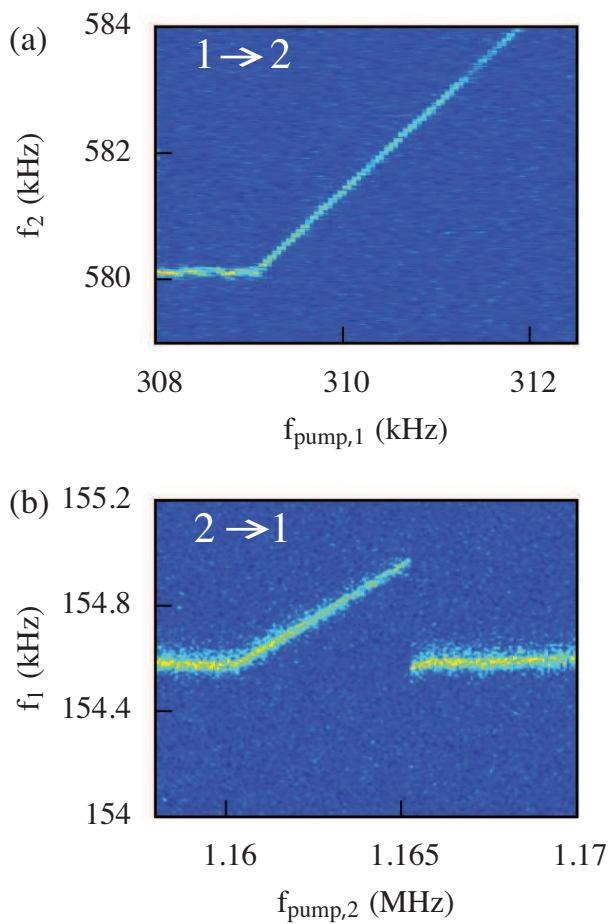}
\caption{Interactions between a directly and parametrically driven mode. (a) Frequency responses of the second mode while varying pump frequency of the first mode. Color scale indicates the amplitude of the second mode. The linear dependence of $f_\mathrm{R,1}$ on $f_\mathrm{pump,2}$ is observed as explained in the text. (b) Reversed experiment; frequency responses of the first mode for varying the pump frequency of the second mode.}
\label{fig4}
\end{figure}

We now investigate the interactions between the different vibrational modes of the same mechanical resonator, when one of the modes is parametrically oscillating. This requires to monitor the response of one mode while another mode is parametrically excited. In particular, the modal interactions between the first and second mode are considered. First, we study the effect of the parametric oscillations of the first mode, characterized in the previous section, on the resonance frequency of the second mode. Fig.~\ref{fig4}(a) shows frequency responses of the second mode, when the first mode is parametrically pumped around its resonance frequency. The first mode is only parametrically pumped and no direct drive at the resonance frequency is applied. Below the resonance frequency of the first mode, no change in resonance frequency of the second mode is observed. Pumping at twice the resonance frequency, the first mode starts to oscillate parametrically. This oscillation induces a significant shift in resonance frequency of the second mode. By parametrically exciting the first mode, the resonance frequency of the second mode is tuned over more than 200 times the resonator linewidth. There is a linear relation of $f_{\mathrm{R,2}}$ on $f_{\mathrm{pump,1}}$ with sensitivity $f_{\mathrm{R,2}}/f_{\mathrm{pump,1}}=1.4$ Hz/Hz.\\
\indent \indent The change in resonance frequency is explained as follows: the oscillation of the first mode increases the tension in the beam. This tension tunes the resonance frequency of the second mode to a higher value. A linear dependence between the two frequencies is expected, as in clamped-clamped beams the resonance frequency of one mode depends quadratically on the amplitude of the other mode~\cite{Westra:2010p7074}, i.e. $f_{\mathrm{R,}i} \sim |A_j|^2$ for modes $i$ and $j$. The amplitude of the parametric oscillation depends on the square root of the pump frequency $|A_j| \sim \sqrt{f_\mathrm{pump,j}}$~\cite{LifshitzCross}, as experimentally verified in Fig.~\ref{fig3}(c). Combining these two dependencies, one expects $f_{\mathrm{R,}i} \sim f_{\mathrm{pump,}j}$. This linear dependence is clearly observed in the measurements, see Fig.~\ref{fig4}(a).   \\

\indent \indent We have also studied the influence of the parametrically excited second mode on the resonance frequency of the first mode, i.e., the first mode is now probing the second mode, which is parametrically oscillating. Again, a linear dependence of the resonance frequency on the parametric pump frequency is found, as is shown in Fig.~\ref{fig4}(b). In this case, the sensitivity $f_{\mathrm{R,1}}/f_{\mathrm{pump,2}}=79$ mHz/Hz. As the pump frequency $f_\mathrm{pump,2}$ is increased above 1.165 MHz the parametric oscillation disappears, and the resonance frequency of the first mode jumps back to its original value. At this point, the nonlinearity causes the oscillation of the second mode to jump to the low amplitude state, which is reflected by the sharp transition of the resonance frequency of the first mode. The large difference in sensitivity with the reversed experiment in Fig.~\ref{fig4}(a) indicates that parametric pumping of the second mode is less effective to change the resonance frequency of the first mode than vice versa. This can be understood since the first mode has the largest oscillation amplitude and can provide the largest tension in the beam. \\

\section{conclusion}
The interactions between a directly and a parametrically oscillating mode of the same mechanical resonator are studied. The parametric amplification and oscillations of a clamped-clamped resonator with an integrated piezoelectric transducer are investigated in detail. The dependence of the oscillation amplitude on pump frequency and phase difference are in agreement with theory. In this work, we demonstrate that the parametric oscillation of one mode induces a change in the resonance frequency of the other vibrational modes. This frequency change is proportional to the pump frequency, as is shown for the first and second mode. The sensitivity of the resonance shift of the second mode on the pump frequency of the first mode is found to be 1.4 Hz/Hz. When the experiment is reversed, i.e. the oscillating second mode is detected by a shift in resonance frequency of the first mode, the sensitivity is 79 mHz/Hz.

\acknowledgements{The authors acknowledge financial support from the Dutch funding organizations FOM (Program 10, Physics for Technology) and NanoNextNL.}


\begin{thebibliography}{26}
\bibitem[{foo()}]{footnote}
\bibinfo{note}{During the experiments, the drift of the resonance frequency is
  within 3 percent.}
\end{thebibliography}
\end{document}